\documentclass[11pt,epsf,amssymb,ulem,qsymbols]{article}

\usepackage{jheppub}

\usepackage{color}
\usepackage{amsmath}
\usepackage{amsmath}
\usepackage{amsfonts}
\usepackage{amssymb}
\usepackage{graphicx}
\usepackage{color}
\usepackage{phonetic}
\usepackage{bbding}
\usepackage{multirow}
\usepackage{rotating}
\usepackage{cprotect}
\usepackage{tablefootnote}
\usepackage{slashed}

\usepackage{bbm}                                                  

\newcommand{\nc}{\newcommand}
\nc{\beq}{\begin{equation}}  \nc{\eeq}{\end{equation}}
\nc{\bea}{\begin{eqnarray}}  \nc{\eea}{\end{eqnarray}}
\nc{\baa}{\begin{array}}     \nc{\eaa}{\end{array}}
\nc{\bit}{\begin{itemize}}   \nc{\eit}{\end{itemize}}
\nc{\ben}{\begin{enumerate}} \nc{\een}{\end{enumerate}}
\nc{\bce}{\begin{center}}    \nc{\ece}{\end{center}}
\nc{\bpm}{\begin{pmatrix}}   \nc{\epm}{\end{pmatrix}}
\nc{\bvt}{\begin{verbatim}}  \nc{\evt}{\end{verbatim}}
\def\half{\frac12}	

\def\to{\rightarrow}

\def\boldoverdot{\,{\raise6pt\hbox{\bf.}\!\!\!\!\>}}
\def\re{{\bf Re}}

\def\acal{{\cal A}}
\def\bcal{{\cal B}}

\def\dcal{{\cal D}}

\def\fcal{{\cal F}}
\def\gcal{{\cal G}}

\def\lcal{{\cal L}}

\def\ncal{{\cal N}}
\def\ocal{{\cal O}}

\def\rcal{{\cal R}}
\def\scal{{\cal S}}

\def\zBB{{\mathbbm Z}}
\usepackage{bm}
\def\diag{\hbox{\diag}}
\def\sm{Standard Model}

\def\gev{\hbox{GeV}}
\def\tev{\hbox{TeV}}

\def\vevof#1{\left\langle #1 \right\rangle}

\def\doubleundertext#1{
{\undertext{\vphantom{y}#1}}\par\nobreak\vskip-\the\baselineskip\vskip4pt%
\undertext{\hbox to 2in{}}}
\def\inbox#1{\vbox{\hrule\hbox{\vrule\kern5pt
     \vbox{\kern5pt#1\kern5pt}\kern5pt\vrule}\hrule}}
\def\sqr#1#2{{\vcenter{\hrule height.#2pt
      \hbox{\vrule width.#2pt height#1pt \kern#1pt
         \vrule width.#2pt}
      \hrule height.#2pt}}}
\def\square{\mathchoice\sqr56\sqr56\sqr{2.1}3\sqr{1.5}3}
\def\today{\ifcase\month\or
  January\or February\or March\or April\or May\or June\or
  July\or August\or September\or October\or November\or December\fi
  \space\number\day, \number\year}
\def\pmb#1{\setbox0=\hbox{#1}%
  \kern-.025em\copy0\kern-\wd0
  \kern.05em\copy0\kern-\wd0
  \kern-.025em\raise.0433em\box0 }

\def\pmbb#1{\setbox0=\hbox{#1}%
  \kern-.02em\copy0\kern-\wd0
  \kern.04em\copy0\kern-\wd0
  \kern-.02em\raise.03464em\box0 }
\def\up#1{^{\left( #1 \right) }}
\def\inv#1{\frac1{#1}}
\def\su#1{{SU(#1)}}
\def\ui{U(1)}
\def\sumprime_#1{\setbox0=\hbox{$\scriptstyle{#1}$}
  \setbox2=\hbox{$\displaystyle{\sum}$}
  \setbox4=\hbox{${}'\mathsurround=0pt$}
  \dimen0=.5\wd0 \advance\dimen0 by-.5\wd2
  \ifdim\dimen0>0pt
  \ifdim\dimen0>\wd4 \kern\wd4 \else\kern\dimen0\fi\fi
\mathop{{\sum}'}_{\kern-\wd4 #1}}
\def\mn{{\mu\nu}}

\def\sm{\scal}
\def\fm{\fcal}

\def\med{{\hbox{\yogh}}}

\def\phit{\tilde\phi}
\def\phitd{\tilde\phi^\dagger}

\def\gdm{\gcal_{\rm DM}}
\def\gsm{\gcal_{\rm SM}}

\def\msc{m_\Phi}
\def\mfe{m_\Psi}

\def\cw{c_{\rm w}}
\def\sw{s_{\rm w}}

\def\ci{c_{\hbox{\tiny\rm I}}}
\def\cii{c_{\hbox{\tiny\rm II}}}
\def\ciii{c_{\hbox{\tiny\rm III}}}

\def\cv{c_{\hbox{\tiny\rm V}}}

\def\cvii{c_{\hbox{\tiny\rm VII}}}

\title{\boldmath Effective theories for Dark Matter interactions and the neutrino portal paradigm. }
\author{Vannia Gonz\'alez Mac\'\i as,~}
\author{Jos\'e Wudka}
\affiliation{Department of  Physics {\it\&} Astronomy,
University of California  Riverside\\ Riverside California 92521-0413, USA}
\emailAdd{vannia.gonzalez-macias@ucr.edu}
\emailAdd{jose.wudka@ucr.edu}

\abstract{
In this article we discuss a general effective-theory description of a multi-component dark sector with an unspecified non-trivial symmetry and its interactions with the Standard Model generated by the exchange of heavy mediators. We then categorize the relevant effective operators given the current experimental sensistivity where the underlying theory is weakly coupled and renormalizable, with neutral mediators: either scalars or fermions. An interesting scenario resulting from this categorization is the neutrino portal, where only fermion mediators are present, and where the dark sector consists of fermions and scalars such that the lightest dark particle is a fermion, this scenario is characterized by having naturally suppressed couplings of the DM to all SM particles except the neutrinos and has received little attention in the literature. 
 We find that there is a wide region in parameter space allowed by the current experimental constraints (relic abundance, direct and indirect detection limits); the cleanest signature of this paradigm is the presence of monochromatic neutrino lines with energy equal to that of the DM mass, but experimental sensitivity would have to be improved significantly before this can be probed. 
}


\begin{document}
\maketitle
\flushbottom

\section{Introduction}

Understanding the fundamental nature of Dark Matter (DM) is one of the most compelling problems in particle physics and cosmology \cite{Feng:2010gw}, yet despite significant and continuing experimental efforts, no information about the nature of DM has been obtained (excepting its gravitational effects \cite{Tegmark,Perlmutter,Tonry,Riess}). In the current favored paradigm  DM is composed of one or more particles whose interactions with the Standard Model (SM) are weak enough to meet the constraints of direct and indirect detection experiments \cite{Akerib:2013tjd,Aprile:2012nq,Choi:2015ara,Aartsen:2012kia}, but strong enough to generate the relic abundance inferred from measurements of the cosmic background radiation \cite{Ade:2013zuv}.

The absence of information on the interactions of the dark sector with the SM, indicates that a model-independent study of these interactions will be useful in understanding the effects of the various possible couplings, and of the processes that generate them; (see, for example, \cite{Goodman:2010ku,Belanger:2008sj,Duch:2014xda}). We will follow such an approach by using an effective Lagrangian  to parametrize the interactions of the dark sector with the SM, and determine the restrictions imposed by the above constraints. We will consider a general dark sector that can include vector, Dirac fermions and scalar particles, and these interact with the SM particles through the exchange of heavy mediators that we denote generically by $\med$~\footnote{In this context, `heavy' indicates that the mediator masses $ M_\med$ are assumed much larger than the typical energies of processes involving interactions between the standard and dark sectors. The standard and dark sectors presumably also interact via the exchange of gravitons; we assume the effects of such interactions are small.}. We will assume that the mediators are weakly coupled to both the standard and dark sectors and that they satisfy the requirements of the decoupling theorem \cite{Appelquist:1974tg}, in particular the interactions they generate between the dark and standard sectors vanish as the mediator mass $ M_\med \to \infty $. 

We will assume that all dark fields transform non-trivially under a symmetry group $ \gdm $ (whose nature we will not need to specify), while all SM particles are assumed to be $ \gdm $ singlets; these characteristics provide a simple way to ensure the dark sector contains a stable particle that will play the role of DM.  Finally, we assume that all dark fields are singlets under the SM gauge group $ \gsm = \su3 \times \su2 \times \ui $.

This paper is organized as follows. In the following section we construct the leading terms in the effective interactions between the standard and dark sectors, we discuss briefly the hierarchy of these couplings generated by the canonical dimension of the corresponding effective operator and by whether it is generated at tree-level or not; we use this to identify the most phenomenologically interesting interactions. We then specialize to the case where the mediators are scalars or fermions and are singlets under $ \gdm \times \gsm $, which we consider separately in sections \ref{sec:scalar.meds} and \ref{sec:fermion.meds}. We then concentrate on a specific scenario, which has received limited attention in the literature: that of a multicomponent dark sector, with fermionic DM and fermionic mediators (section \ref{sec:fermionic.DM}). Parting comments are provided in section \ref{sec:conclusions} while details of the calculations are relegated to the appendices.

\section{DM-SM interactions}
An immediate consequence of our assuming mediator-generated interactions between the  standard and dark sector is that these interactinos take the form 
\beq
\ocal = \ocal_{\rm SM} \ocal_{\rm dark} 
\label{eq:generic}
\eeq
where $ \ocal_{\rm SM,\, dark} $ denote local operators composed of standard and dark fields  respectively, and which are invariant under both $ \gdm$ and $ \gsm $, but need {\em not} be Lorentz singlets. The operators $\ocal$ will apepar in the effective Lagrangian multiplied by coefficients proportional to $ 1/M_\med^n $ with $ n = {\rm dim}(\ocal)-4 $; in particular, the larger the dimension of $ \ocal $, the smaller its effect, so that within this paradigm, dominant effects will be generated by lower dimensional operators. Given the detailed knowledge of the SM constructing the operators $ \ocal_{\rm SM} $ invariant under $ \gsm \times \gdm$ is a straightforward exercise~\footnote{Invariance under $ \gdm $ is automatic since all SM fields are assumed to be $ \gdm$ singlets.}. 

When constructing $ \ocal_{\rm dark} $ we will assume that the dark sector is composed of scalars $ \Phi $, Dirac fermions $ \Psi $ and vectors $X$, with the understanding that the dark sector present in Nature may only contain a subset of these particles. Again by assumption all dark fields are invariant under $ \gsm $, so that $ \ocal_{\rm dark} $ will also be a singlet under this group. Whether a given combination of dark fields is invariant under $ \gdm $ or not depends on the details of this symmetry and the representations carried by these fields, unfortunately existing data does not provide any information on this point, so we opt for the most general case and study the effects of all interactions of the form (\ref{eq:generic}) when $ \ocal_{\rm dark} $ is invariant under at least one choice of $ \gdm $; we can only say that since all dark fields are assumed to transform non-trivially under $ \gdm $, $ \ocal_{\rm dark} $ will contain at least two fields. The list of the operators $ \ocal $ of dimension $ \le 6$ that satisfy the above conditions is given in table \ref{tab:ops}.

\begin{table}[t]
$$
\begin{array}{cc|c}
{\rm dim.} 			& {\rm category} &\cr\hline
4 					& {\rm I} & |\phi|^2 (\Phi^\dagger \Phi)														\cr \hline
\multirow{3}{*}{5}	& {\rm II} & |\phi|^2 \bar\Psi \Psi \qquad |\phi|^2 \Phi^3 									\cr
					& {\rm III} & (\bar\Psi \Phi)(\phi^T \epsilon \ell) 													\cr
					& {\rm IV} & B_\mn X^\mn \Phi \qquad B_\mn \bar\Psi \sigma^\mn \Psi					\cr\hline
\multirow{4}{*}{6}	& {\rm V} & |\phi|^2 \ocal_{\rm dark}\up4 \quad \Phi^2 \ocal_{\rm SM}\up4 					\cr
					& {\rm VI} & (\bar\Psi \Phi^2)(\phi^T \epsilon \ell) \quad (\bar\Psi \Phi)\slashed{\partial} (\phi^T \epsilon \ell)	\cr
					& {\rm VII} & J_{\rm SM}.J_{\rm dark}															\cr
					& {\rm VIII} & B_\mn \ocal_{\rm dark}^{(4)\,\mn} \cr\hline
\end{array}
$$
\caption{\small List of operators of dimension $\le6$ involving dark and SM fields; $ \phi $  denotes the SM scalar isodoublet, $B$ the hypercharge gauge field, and $\ell$ a left-handed lepton isodoublet;  also, $ \epsilon= i \sigma_2$, where $ \sigma_2$ denotes the usual Pauli matrix. Dark scalars, Dirac dark fermions and vectors are denoted by $ \Phi,\, \Psi$ and $X$ respectively. The currents in category VII operators are defined in (\ref{eq:currents}), and the operators $ \ocal\up4$ in categories V and VIII are listed in appendix \ref{sec:o4}. See the text for details.}
\label{tab:ops}
\end{table}

\medskip Referring to this list we make the following observations:
\bit
\item Some of the operators might be disallowed by the spectrum in the dark sector (e.g. the operator in category I would be absent if there are no dark scalars). Other operators are present only when the dark sector has several components (e.g. dark fermions and scalars are required for the operator in category III to be present).

\item Though it is customary to assume $ \gdm $ is a discrete symmetry (e.g. $ \zBB_n $) this need not be the case. For example, $\gdm$ could be a gauge group, and $X$ the associated non-Abelian (dark) gauge bosons. In this case operators such as $B_\mn X^\mn \Phi$ would be invariant if the corresponding $ \Phi $ belong to the adjoint representation of $ \gdm $.

\item Some operators might be disallowed by the choice of $ \gdm $; for example, if $ \gdm = \zBB_2 $, our requirement that all fields transform non-trivially under this symmetry requires them all to be odd under $ \zBB_2 $, hence $ |\phi|^2 \Phi ^3$ is forbidden.
 
\item The $ \ocal\up4$ in categories V and VIII represent dimension-4 field combinations of the corresponding sector, invariant under $\gdm$ and $\gsm$; they are listed in appendix \ref{sec:o4}. 

\item In category VII, $J_{\rm SM,\,dark}$ represent SM and dark currents of dimension 3, invariant under $ \gdm$ and $ \gsm $:
\beq
\begin{array}{ll}
J_{\rm SM}^{(\psi)\,\mu} = \bar\psi \gamma^\mu \psi\,, & \quad
J_{\rm SM}^{(\phi)\,\mu} =\inv{2i} \phi^\dagger \stackrel\leftrightarrow{D^\mu} \phi\,, \cr
J_{\rm dark}^{(L,R)\,\mu} = \bar\Psi \gamma^\mu P_{L,R}\Psi\, & \quad
J_{\rm dark}^{(\Phi)\,\mu} = \inv{2i} \Phi^\dagger \stackrel\leftrightarrow{\dcal^\mu} \Phi
\end{array}
\label{eq:currents}
\eeq
where $ \psi $ denotes any SM fermion, $D$ the covariant derivative in the standard sector, and $ \dcal $  the covariant derivative in the dark sector (replaced by an ordinary derivative if this sector is not gauged).

\item In general there can be more than one dark field of each type, but we did not include such `dark-flavor' indices for notational simplicity; similarly we did not include a generation index in the left-handed lepton isodoublet $ \ell $ (categories III and VI).

\item There are no operators of the form $ X_\mn \ocal_{\rm SM}^{(4)\,\mn}$ because they would not invariant under $ \gdm$, given our assumption that all dark fields transform non trivially under this group.
\eit

\subsection{Tree-level and loop-level generated operators -- neutral mediators}

All the operators in table \ref{tab:ops} will, in general, be generated in any model within the class of theories being considered here. Depending on the details of the model, any given operator will be generated at tree level, or at one (or higher loops) involving a combination of mediator, dark and standard internal lines~\footnote{Our effective theory will be valid at scales below the mediator mass $ M_\med$, and is generated by integrating out the mediators as well as all standard and dark field modes with momenta above $ M_\med $.}. Operators that are generated at $L$-loops will have coefficients suppressed~\footnote{The presence of loop-suppression factors does not necessarily mean that the effects of the corresponding operators are phenomenologically irrelevant since the experimental sensitivity that may be sufficiently high to warrant retaining them; ignoring such factors, however, can lead to significant {\em over}-estimation of these effects.} by a factor $ \sim 1/(4\pi)^{2L} $; tree-generated operators will have no such suppression. Because of this, evaluating the effects of an operator and eliciting constraints from data on the model parameters will depend strongly on whether the operator is tree or loop generated, and this requires that some properties of the mediator sector be specified. 

We will now restrict ourselves to models where the dark and a standard sector interact through the exchange of neutral mediators, that is, which are singlets under $ \gdm \times \gsm $. We will also assume that the full theory composed of mediators, dark and standard sector is renormalizable and that the mediators are weakly coupled. Within this neutral-mediator paradigm one can determine by inspection that the operators in table \ref{tab:ops} are generated at tree level by scalar mediators $\sm$ (categories II an V), fermion mediators $\fm$ (categories III and VI), vector mediators (category VII), or antisymmetric tensor mediators (categories IV and VIII). In this paper we will consider only the case of scalar and fermion mediators; tensor and vector mediators require a discussion of the corresponding local symmetries and will be presented elsewhere. The operators that can be generated at tree-level by scalar and fermion neutral mediators are listed in table \ref{tab:ptg.list}

\begin{table}
$$
\begin{array}{c|l|l}
\med	& \multicolumn{1}{c|}{\rm dim=5}			& \multicolumn{1}{c}{\rm dim=6}								\cr
\hline
\sm	& |\phi|^2 \bar\Psi \Psi, \quad |\phi|^2 \Phi^3	& |\phi|^2 \Phi^4, \quad \Phi^2 |\phi|^4							\cr
\fm	& (\bar\Psi \Phi)(\phitd \ell)					& (\bar\Psi \Phi^2)(\phitd\ell), \quad (\bar\Psi \Phi)\slashed{\partial} (\phitd \ell)	\cr
\end{array}
$$
\caption{Dark-SM operators that are generated at tree level by neutral (under $ \gdm$ and $ \gsm $) scalar ($\sm$) and fermion ($\fm$) mediators. Some of the operators in category V (see table \ref{tab:ops}) are not listed because they are generated at tree-level only when heavy charged (under the dark or SM symmetries) particles are present. The currents $ J_{\rm SM,\, dark}$ are assumed to be conserved (see text).}
\label{tab:ptg.list}
\end{table}

\subsection{The effective Lagrangian}
For processes with typical energies below the mediator mass the relevant physics is described by an effective theory resulting from integrating out all modes with energies above this scale; aside from the mediators themselves, these modes include those of the standard model and dark fields with momenta above $M_\med$. In the following we refer to these as `high-momentum modes' (HMM). Integrating out the $ \med $ and HMM generates an effective Lagrangian of the form
\beq
\lcal_{\rm eff}\up\med = \lcal_{\rm SM} + \lcal_{\rm dark} + \ci |\phi|^2 |\Phi|^2 + \lcal\up{\med{\rm-tree}} + \lcal\up{\med{\rm-loop}}
\label{eq:leff}
\eeq
where the first and second terms correspond to the Lagrangians for the standard and dark sectors (including the loop-generated interactions resulting from integrating out the HMM within the corresponding sector). The third term contains the renormalizable Higgs-portal interaction(s) in category I, where $ \ci \sim 1 $ is the natural value of this coefficient (this term is of course absent if the dark sector does not contain scalars). 

The fourth term contains the tree-level generated effective dark-SM interactions resulting form integrating out the mediators only; these are listed in table \ref{tab:ops} . The last term contains all loop-generated terms obtained by integrating out mediators and HMM, they generate the remaining terms in table \ref{tab:ops} but the corresponding coefficients suppressed by loop factors $ \sim 1/(4\pi)^{2L} $, where $L$ is the number of loops. The interactions contained in these last two terms will be suppressed by inverse powers of the mediator mass~\footnote{$M_\med$ is the same scale used in integrating-out the HMM.} scale $M_\med $.

\subsubsection{Some simplifications}
\label{sec:simpl}

For the case where the SM and dark sectors are connected only by exchange of neutral mediators the effects of several operators in table \ref{tab:ops} are either very small effects or subdominant. In the following we will ignore such contributions to the effective Lagrangian (recognizing that in near-future experiments they might generate small but potentially observable deviations from the leading effects). The operators that we will drop are~\footnote{These considerations depend strongly on our assumptions of weak coupling and renormalizability.}:

\bit
\item{\sl Operators of dimension $\ge5 $ with only scalar fields.} Given the current experimental sensitivity, the effects of operators in categories II and V that involve only scalars ($\phi$ and $ \Phi $; see also appendix \ref{sec:o4}) will be subdominant compared to those generated by the Higgs portal coupling $ \propto \ci$ in (\ref{eq:leff}), provided all coefficients are within their natural ranges. 

\item{\sl Operators in category VI.} The observable effects generated by these operators are very similar but subdominant to those generated by the operator in category III. 

\item{\sl Operators generated at $\ge2$ loops.} Such operators appear multiplied by a small coefficient $ \sim1/(4\pi)^4 \sim 4 \times 10^{-5}$. Specifically, these are: 
\bit
\item Category III operators when only $\sm$ mediators are present.
\item All category IV and VIII operators.
\item The operators $|\Phi|^2 (G_\mn^A)^2$ in category V.
\item Category VII operators when only $\sm$ mediators are present, or when $ \fm$ mediators occur and the operators involve $J\up i_{\rm SM} $ for $ i \not=\ell,\, \phi$.
\eit
\eit
The remaining operators are listed in table \ref{tab:rem-ops}.

\begin{table}[t]
$$
\begin{array}{cc|c}
{\rm dim.} 			& {\rm category} &\cr\hline
\multirow{2}{*}{5}	& {\rm II}	& |\phi|^2 \bar\Psi \Psi   																					\cr
					& {\rm III}	& (\bar\Psi \Phi)(\phi^T \epsilon \ell) 																		\cr \hline
\multirow{2}{*}{6}	& {\rm V}	&  |\phi|^2 \bar\Psi \Phi \Psi',~ |\phi|^2 X_\mn^2,~  \Phi^2 \bar\psi \varphi \psi',~ \Phi^2 B_\mn^2,~ \Phi^2 (W_\mn^I)^2	\cr
					& {\rm VII}	& J_{\rm SM}\up i  \cdot J_{\rm dark}\up a ~ (i=\ell,\phi;~a=\Phi,L,R)
\end{array}
$$
\caption{\small Leading dark-SM interactions  of dimensions $ 5 $ and $6$ (see section \ref{sec:simpl}); the operators in categories III and VII should be removed when only scalar mediators are present.}
\label{tab:rem-ops}
\end{table}

\section{Scalar mediators}
\label{sec:scalar.meds}
When scalar mediators are present, the list of tree-generated operators is on the first line of table \ref{tab:ptg.list}; of the remaining operators in table \ref{tab:ops} we retain only those generated at one loop. The resulting effective Lagrangian is
\bea
\lcal\up{\sm-{\rm treee}} &=&\frac{\cii}\Lambda |\phi|^2 \bar\Psi \Psi + \cdots \cr
\lcal\up{\sm-{\rm loop}} &=&  \sumprime_r  \frac{\cv^r}{(4\pi \Lambda)^2}\ocal\up6_r \cdots
\label{eq:leffs}
\eea
where the prime indicates that the sum is over the category V operators listed in table \ref{tab:rem-ops}.

In the case where both dark scalars and fermions are present the $\sm$ will also generate the dimension 5 interaction $ |\Phi|^2( \bar\Psi \Psi) $ so that, only the lightest of these particles will contribute significantly to the relic density (unless the mass splitting is small or some coefficients are significantly suppressed  \cite{Drozd:2011aa,Bhattacharya:2013hva,Kumar:2013tra}). It then follows that, as far as direct and indirect detection are concerned one can consider a single-component model where only the lightest of the dark particles is included. Such single-component DM models are obtained by retaining the appropriate subset of the above interactions. In particular, the Higgs-portal coupling (proportional to $\ci $) is the most significant interaction in models where only dark scalars are present; this type of models have been extensively studied in the literature \cite{Silveira:1985rk,McDonald:1993ex,Burgess:2000yq,Bento:2000ah,Holz:2001cb,LopezHonorez:2006gr} .

\section{Fermion mediators}
\label{sec:fermion.meds}
When only fermion mediators are present~\footnote{When both scalar and fermion neutral mediators are present the effective theory is obtained by adding (\ref{eq:leffs}) and (\ref{eq:lefff}). The resulting effective theory has an $O(1) $ coupling for the operators in categories II and III, and $O(1/16\pi^2)$ for categories V and VII.} the effective Lagrangian resting form integrating the $\fm $ takes the form
\bea
\lcal\up{\fm{\rm-tree}} &=& \frac{\ciii}\Lambda (\bar\Psi \Phi)(\phitd \ell) + \cdots \cr
\lcal\up{\fm{\rm-loop}} &=& \frac{\cii}{16\pi^2 \Lambda}|\phi|^2 \bar\Psi \Psi + \sum_{a=\ell\,\phi;\, i=L,R,\Phi} 
\frac{\cvii\up{a | i}}{(4\pi \Lambda)^2} \left( J\up a_{\rm SM} \cdot J\up i_{\rm dark} \right)
+ \sumprime_r \frac{\cv^r}{(4\pi \Lambda)^2} \ocal\up6_r + \cdots 
 \cr &&
\label{eq:lefff}
\eea
where the prime indicates that the sum is over the category V operators listed in table \ref{tab:rem-ops}, and the ellipses denote subdominant operators (see section \ref{sec:simpl}); the currents are defined in (\ref{eq:currents}). It is also worth noting that the category II operator $ |\phi|^2 \bar\Psi \Psi $ has a loop-suppressed coefficient for these models, which was not the case when scalar mediators are present.

When expanded in terms of component fields in the unitary gauge (for the SM) the dimension 5 term in (\ref{eq:lefff}) is seen to contain a vertex of the form $ \ciii \bar\nu_L \Phi^\dagger \Psi $, where $ \nu_L $ denotes a SM left-handed neutrino. The presence of this coupling implies that the heavier of the dark particles $(\Phi, \Psi) $ will decay promptly into the lighter one and a neutrino. So this type of models, while having a multi-component dark sector, have a single-component DM relic~\footnote{Since the neutrino masses are so small there is no need to assume a large splitting between $\msc$ and $ \mfe$ to differentiate between these two scenarios. The more complicated case where the dark fermions and scalars are degenerate is unnatural as there is no symmetry that can ensure this constraint. For example, if the dark sector is assumed to be supersymmetric with $ \Phi $ and $ \Psi $ members of a multiplet, then the category III operator explicitly breaks this dark supersymmetry.}. 

If $ \msc < \mfe $ the relic DM is composed solely of scalars $ \Phi $ and the model largely reduces to the one discussed in the simplest Higgs-portal models (when some of the coefficients are suppressed by $ (4\pi)^2 $).

If, however, $ \mfe < \msc $, the situation is different: DM will be composed of the dark fermions $ \Psi $, and the leading $ \Psi $ interactions are with neutrinos or neutrinos and Higgs particles. This is a case that we discuss in more detail below, and which has not received much attention in the literature (right-handed neutrino portals with lepton number violation were examined, for example in \cite{Falkowski:2011xh,Baek:2013qwa}; and models including heavy Majorana right-handed neutrino mediators and fermion DM relics were proposed to explain dark matter-antimatter asymmetry through leptogenesis \cite{Cosme:2005sb,An:2009vq,Falkowski:2011xh}).

\section{Fermion DM with fermion mediators: neutrino portal DM}
\label{sec:fermionic.DM}
We now consider in some detail the case where the dark sector contains Dirac fermions and scalars, with the latter being heavier than the former, and when dark particles interact with the SM via the exchange of neutral Dirac fermions conserving lepton number~\footnote{When  model-building, small violations of lepton number can be introduced in the mediator sector as perturbations. This would not affect the discussion below in any significant (qualitative and quantitative) way.} . As indicated above, the main DM-SM interactions involve neutrinos or Higgs particles, while all other interactions have loop-suppressed couplings.

\subsection{Relic abundance }
\label{relic}

The leading DM-SM interaction is generated by the (tree-level) exchange of the dark scalars $ \Phi $ and represent the most important reaction responsible for the equilibration between the dark and standard sectors in the early universe; this process is produced by the interaction $ \propto \ciii $ in (\ref{eq:lefff}). It is worth noting that even if the dark scalars do not contribute to the relic abundance, their presence is essential for the viability of the model: in the absence of $ \Phi $ all terms in (\ref{eq:lefff}) would be absent~\footnote{For the loop-suppressed terms this follows from a straightforward examination of the possible diagrams.} and the SM and dark sectors would decouple.

The remaining interactions in $ \lcal\up{\fm-\rm{loop}}$ generate small corrections but for two exceptions: the terms $ \propto \cii $ and $ \propto \cvii\up{\phi | L , R} $ contain the vertices $ \bar\Psi \Psi H$ (where $H$ denotes the Higgs) and $ \bar\Psi \slashed{Z} P_{L,R} \Psi $, and will produce important resonant effects when $ \mfe \simeq m_H/2 $ and $ \mfe \simeq m_Z/2 $ that are generated, respectively, by the $s$-channel exchange of the $H$ and $Z$. Other interactions of the form $ J_{\rm SM} \cdot J_{\rm dark} $ or $ \ocal\up6_r$ in (\ref{eq:lefff}) generate small corrections that we will ignore in the following.

In the unitary gauge
\beq
(\bar\Psi \Phi)(\phitd \ell) \supset \frac v{\sqrt{2}} (\bar\Psi \nu \Phi) \,,\quad
\bar\Psi \Psi |\phi|^2 \supset v H (\bar\Psi \Psi ) + \,,\quad
J_{\rm SM}\up\phi \cdot J_{\rm dark}\up{L,R} \supset - v m_Z \bar\Psi \slashed{Z} P_{L,R} \Psi \,;
\label{eq:unit_int}
\eeq
 with $ v \sim246 \gev $. The cross section for $\Psi \Psi \to \nu \nu$ (relevant for the relic abundance calculation below) is generated by the diagrams in figure \ref{fig:PPnn} and can be obtained using standard techniques; we include the analytic expression in appendix \ref{sec:xs}.

\begin{figure}[th]
$$\includegraphics[width=2 in]{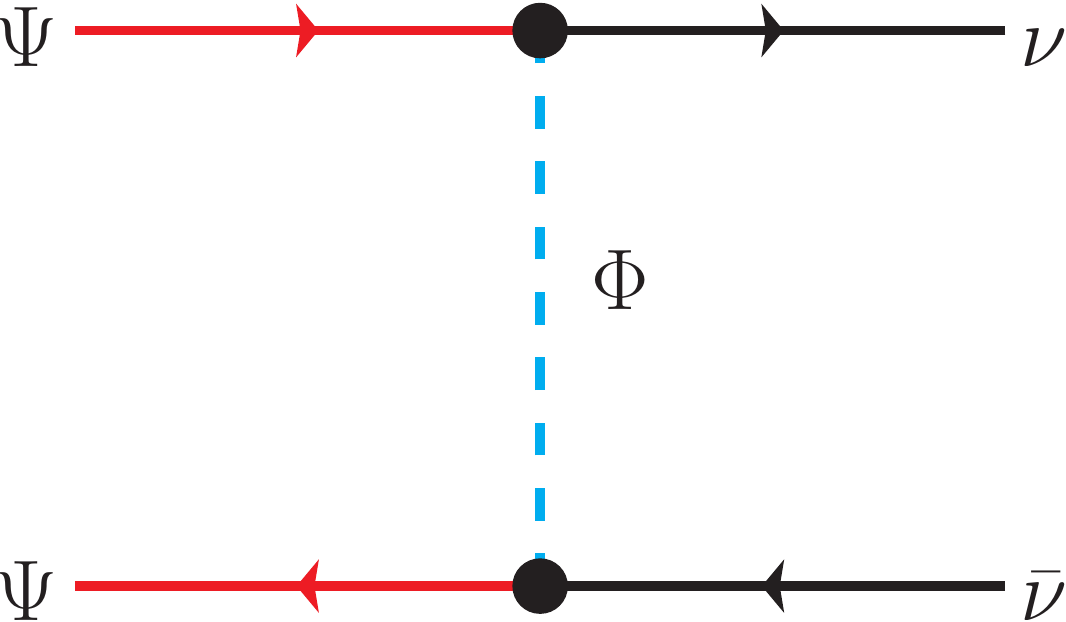} \qquad
\includegraphics[width=2 in]{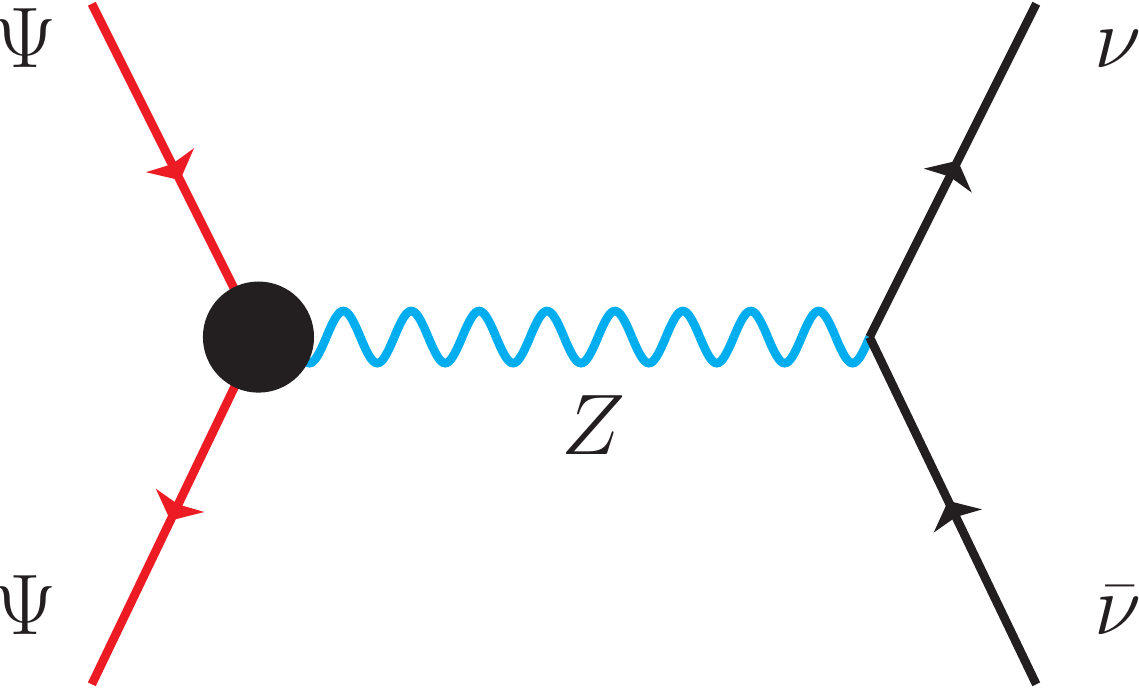}$$
\caption{\small Leading DM-SM interactions for the case where the effective vertices (represented by black circles) are generated by neutral fermionic mediators. The $t$-channel diagram ($\Phi$ exchange) generates the leading contribution $\propto |\ciii|^4$; the $s$-channel diagram ($Z$ exchange) generates a resonant contribution $\propto |\cvii\up{\phi | L,R} |^2 $ that is significant only when $ \mfe \simeq m_Z/2 $; see (\ref{eq:sv1b}).}
\label{fig:PPnn}
\end{figure}

The cross section for $\Psi\Psi$ annihilation into heavier fermions has two resonant contributions from the Higgs and the $Z$ boson generated by the diagrams in figure \ref{fig:PPffH}; their expressions are also given in appendix \ref{sec:xs}. From these results, and using the approximations described in \cite{Kolb:1990vq}, we readily obtain,
%
\bea
\vevof{\sigma v}_{\Psi \Psi \xrightarrow[]{H} ff}&\simeq& \frac{N_fm^2_f}{4\pi \mfe}\left(\frac{\cii }{16\pi^2\Lambda }\right)^2 
\frac{ (\mfe^2-m^2_f )^{3/2}}{(m^2_H-4 \mfe^2)^2+m^2_H\Gamma^2_H} \\
\label{eq:ztt}
%
%
\vevof{\sigma v}_{\Psi \Psi \xrightarrow[]{Z} bb} &\simeq&  \bar\sigma_Z \left[
 \frac{9}{4}(1+4\bcal)-\frac{3}{2}u_b (\bcal- 1)   +  \left(1+4\bcal+2 u_b \bcal \right) \sw^2(2 \sw^2-3)  \right]\\
\vevof{\sigma v}_{\Psi \Psi \xrightarrow[]{Z} \tau\tau} &\simeq& 12 \bar\sigma_Z \left[
  1+4\bcal   -2 u_\tau (\bcal - 1)  +
4 \left(  1+4 \bcal     + 2 u_\tau \bcal \right)  \sw^2(2\sw^2-1)  \right]\\
\label{eq:sv1a}
\vevof{\sigma v}_{\Psi \Psi \to \nu \nu} &\simeq& \frac{(v /\Lambda_{\rm eff})^4}{256\pi\mfe^2}\left[ \left| \half + B_L+B_R \right|^2 + \frac34 \right]
\label{eq:sv1b}
\eea
where in the first line we ignored $O(\mfe^2/\Lambda^2)$ corrections and
\bea
 B_{L,R}&=&\left( 1+\frac{m^2_\Phi}{\mfe^2}\right)
 \left(\frac g{4 \pi c_{\rm w}} \right)^2 \frac{\cvii\up{\phi | L,R}}{\ciii^2} \frac{\mfe^2}{ m_Z^2-4\mfe^2 + i m_Z \Gamma_Z} \cr && \cr
\Lambda_{\rm eff}& =& \sqrt{ 1 + \frac{\msc^2}{\mfe^2} } \frac\Lambda \ciii,\,  \qquad
\bcal = \frac{|B_L +  B_R |^2}{ |B_L|^2 +| B_R |^2},\, \qquad u_i = \frac{m_i^2}{\mfe^2} \cr && \cr
 \bar\sigma_Z &=& \frac{(v /\Lambda_{\rm eff})^4(  |B_L |^2+ |B_R|^2)}{2048\sqrt{3}\pi\mfe^2} 
\label{eq:leff}
\eea
The expressions  (\ref{eq:ztt}-\ref{eq:sv1b})  correspond to the $s$-wave annihilation processes for the corresponding channels.

Using standard results \cite{Kolb:1990vq} we use these expressions to derive the relic abundance:
\beq
\Omega_\Psi h^2 = \frac{1.07 \times 10^9}\gev \, \frac{ x_f }\xi \,; \quad \xi = \frac{ M_{\rm Pl} \vevof{\sigma v}_{\rm tot} }{\sqrt{g_\star}}\,,
~\vevof{\sigma v}_{\rm tot} = \sum_f \vevof{\sigma v}_{\Psi \Psi \xrightarrow[]{Z,H} ff}\,,
\label{eq:Omapprox}
\eeq
where $M_{\rm Pl}$ denotes the Planck mass, $ g_{\star S} ,\, g_\star $ denote, respectively, the relativistic degrees of freedom associated with the entropy and energy density, and
\beq
x_f = \frac\mfe{T_f} = \ln \left(0.152 \mfe \xi \right) - \half \ln \left[ \ln \left(0.152 \mfe \xi \right) \right]\,,
\eeq
and $T_f$ is the freeze-out temperature.

\begin{figure}[th]
$$\includegraphics[width=2in]{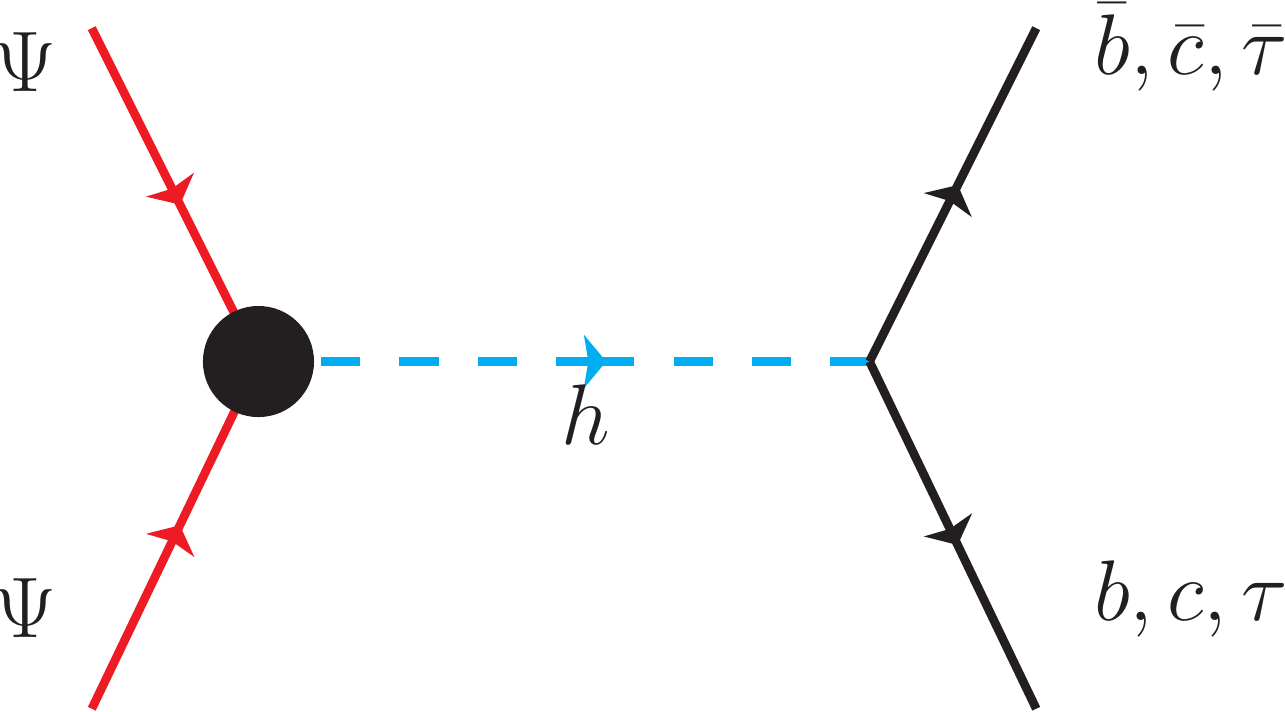} \qquad
\includegraphics[width=2in]{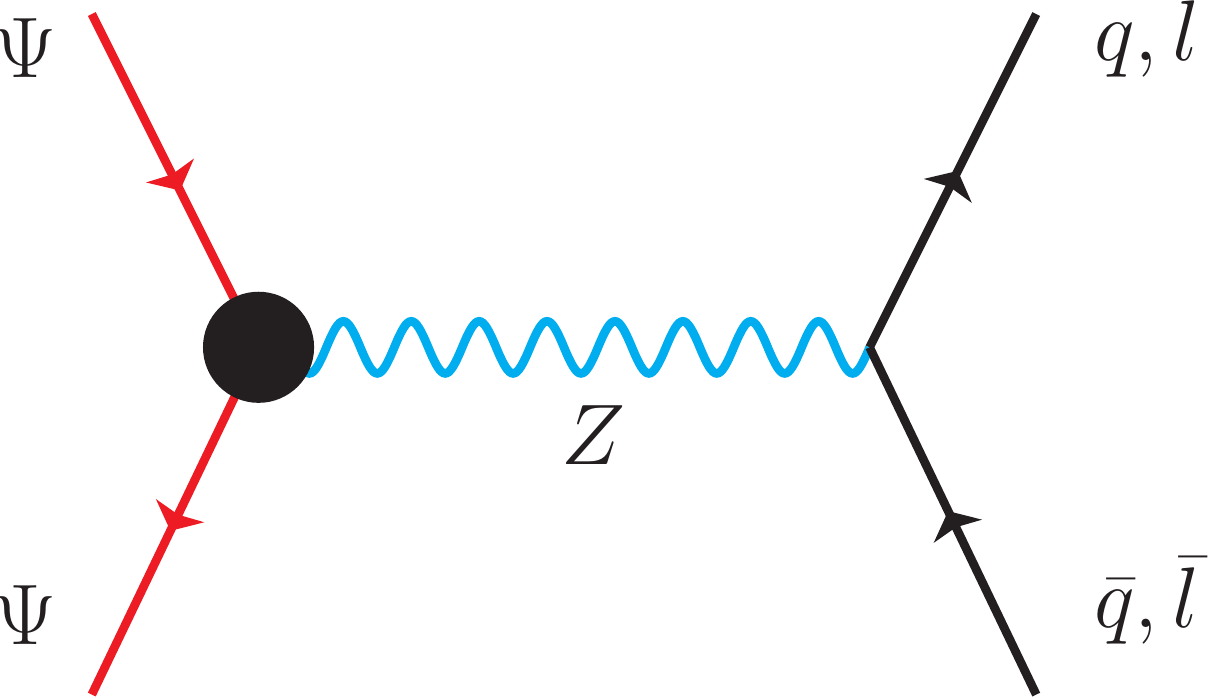} $$
\caption{\small Resonant contributions for the $\Psi\Psi\to ff$ annihilation cross section.}
\label{fig:PPffH}
\end{figure}

\begin{figure}
$$\includegraphics[width=5.in]{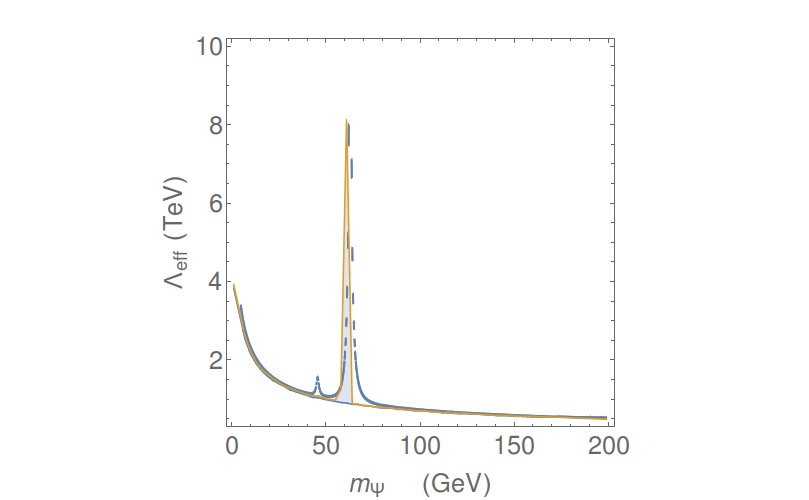}$$
\cprotect\caption{\small Constraints on the fermionic DM model derived from the relic abundance inferred from the Plank data (at $3\sigma$): the light blue region delimited by the yellow line is the allowed region calculated using the \verb|MicrOmegas| code; the thick blue is obtained using (\ref{eq:Omapprox}) with (for illustration purposes) $\cii=0.08,\, \ciii=1.5,\, \cvii\up{\phi|R}=-5,\, \cvii\up{\phi| L}=-5$ (the effective scale $ \Lambda_{\rm eff} $ is defined in (\ref{eq:leff})). The graph clearly exhibits the two resonant peaks at $\mfe=m_Z/2$ and $\mfe=m_H/2$.}
\label{fig:relic_f}
\end{figure}

The expression for $ \Omega $ can now be compared to the result inferred from CMB data obtained by the Planck experiment\cite{Ade:2013zuv}:
\beq
\Omega_{\rm Planck} h^2 = 0.1198\pm0.0026 \quad (3\,\sigma).
\label{eq:planck}
\eeq
Outside the resonance region $ \Omega $ is determined by the $ \Psi \Psi \to \nu\nu $ cross section (\ref{eq:sv1b}) and so will be a function of $ \Lambda_{\rm eff} $ and $ \mfe $; accordingly  (\ref{eq:planck})  selects a narrow region in  the  $(\mfe, \Lambda_{\rm eff})$ plane (see figure \ref{fig:relic_f}), which is well approximated by the relation
\beq
\Lambda_{\rm eff} \simeq \sqrt{\frac{m_\Omega }\mfe }\, \tev; \quad m_\Omega \simeq 74\gev \quad (\hbox{non-resonant~region}).
\label{eq:Leff.approx}
\eeq

In addition to the above analytic calculation, we also derived numerically the constraints on the model parameters. This calculation was done by selecting $ 2 \times 10^7$ points in the 7-dimensional parameter space $ \{\cii,\, \ciii,\, \cvii\up{\phi | L,R},\, \Lambda,\, \mfe,\, m_\Phi \} $ within the ranges
\bea
&1\,\gev	\leq \mfe			\leq 199\, \gev \,, \quad
1\, \tev	\leq \Lambda		\leq 5\, \tev \,, \quad
11\,\gev		\leq \msc			\leq 836\,\gev, & \cr &&\cr
&0 		\leq \ciii 			\leq 4\,, \quad
		 |\cii|			\leq 10\,, \quad
-8		\leq \cvii\up{\phi|L}	\leq 0 \,, \quad
-10		\leq \cvii\up{\phi|R}	\leq 0 \,.&
\label{eq:ranges}
\eea
while keeping $ \mfe < \msc $. For each point $ \Omega $ was obtained using the public codes \verb|MicrOmegas| \cite{Belanger:2013oya} and \verb|CALCHEP| \cite{Belyaev:2012qa} (model implementation for \verb|CALCHEP| was done using the \verb|FeynRules| package \cite{Alloul:2013bka}). The results are also presented in figure \ref{fig:relic_f} together with the comparison to the analytic expressions.  

\subsection{Direct detection}

At present the most stringent limit on spin-independent scattering cross sections of DM-nucleon particles comes from the LUX experiment \cite{Akerib:2013tjd}. In order to derive the implications for the effective theory under study we obtained the DM-nucleon $\Psi \ncal\to \Psi \ncal$ cross sections in the limit where the relative velocity $v\to 0$ (we again use \verb|MicrOmegas|). In this non-relativistic limit the elastic amplitudes are divided into spin-independent interactions (generated by scalar and vector couplings) or spin-dependent interactions (generated by axial-vector couplings) . 

The terms in (\ref{eq:lefff}) that are relevant for direct detection are all contained in $ \lcal\up{\fm{\rm-loop}} $, specifically the terms proportional to $\cii$ and to $\cvii\up{\phi|L,R}$:
\beq
\lcal\up{\fm{\rm-loop}} = \frac{v \cii}{16\pi^2 \Lambda} H \bar\Psi \Psi - \frac g{2 \cw} \frac{v^2 }{16\pi^2 \Lambda^2} \bar\Psi \slashed{Z} \left ( \cvii\up{\phi|L} P_L + \cvii\up{\phi|R} P_R \right)
 \Psi + \cdots 
 \label{ZHex}
\eeq 
where we used (\ref{eq:unit_int}); the first and second terms generate DM-SM interactions via Higgs and $Z$ exchanges, respectively. From this expression we extract
\beq
\epsilon_H = \frac{v^3}{16\pi^2 \Lambda m_H^2}\cii \,; \qquad \epsilon_{Z} = - \frac{ v^2 }{16\pi^2 \Lambda^2} \frac{\cvii\up{\phi|L} + \cvii\up{\phi|R}}2
\label{eq:eps}
\eeq
that provide estimates of the strength of the $H$ and $Z$ exchanges to the total direct-detection cross section. The final result takes the form
\beq
\sigma_{\Psi \ncal\to\Psi \ncal}=\frac4\pi \mu_{\rm red}^2 \left| (Z_{\rm nucl}/A_{\rm nucl}) \acal_p +(1- Z_{\rm nucl}/A_{\rm nucl}) \acal_n \right|^2
\eeq 
where $\acal_{p,n} $ denote, respectively the amplitudes for proton and neutron scattering (in units of $1/$mass$^2$),  $A_{\rm nucl},\, Z_{\rm nucl}$ denote the atomic number and nuclear charge respectively, and $ \mu_{\rm red}$ the $\ncal-\Psi$ reduced mass

\begin{figure}
$$\includegraphics[width=4.5in]{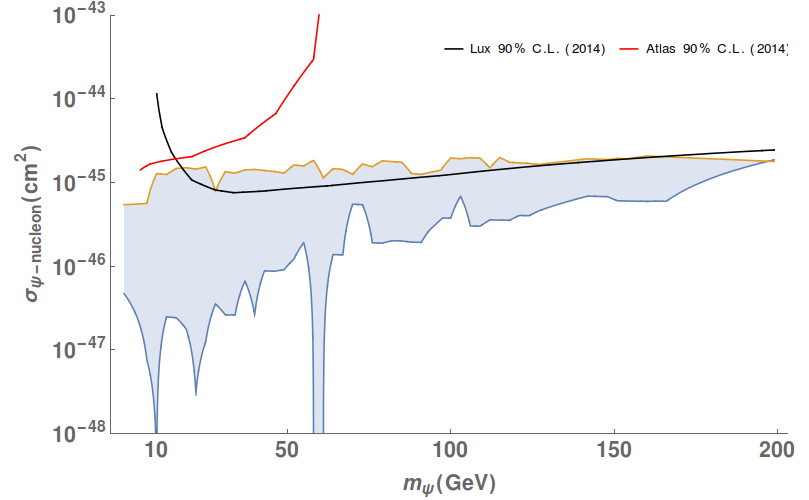}$$
\caption{\small Shaded region: DM-nucleon cross sections for the parameter ranges (\ref{eq:ranges}) of the neutrino-portal DM scenario, all points satisfy the relic abundance constraint (the discontinuous boundary is due to limitations in parameter sampling, except for the Higgs resonance peak at $\mfe \sim 60 \gev $). Black line: limit form the LUX experiment at 90\% C.L. \cite{Akerib:2013tjd}; red line: ATLAS limit \cite{Aad:2014iia} derived from the invisible decay of the Higgs at 90\% C.L.}
\label{fig:SigmaXN}
\end{figure}

\begin{figure}
$$\includegraphics[width=4.5in]{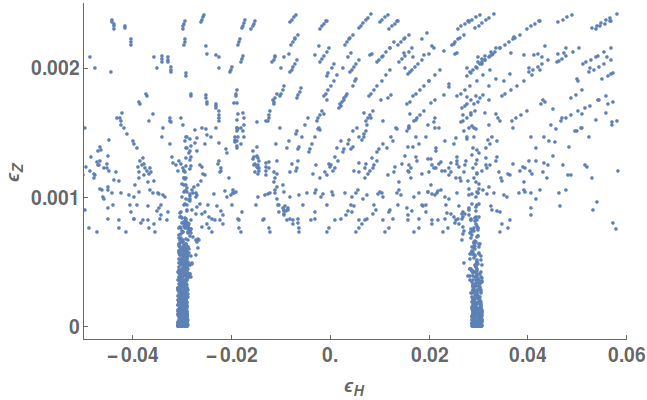}$$
\cprotect\caption{Allowed values of $\epsilon_h$ and $\epsilon_V$ by the LUX experiment (at $90\%$ C.L.) and the PLANCK constraints from the CMB.}
\label{fig:EpsSigma}
\end{figure}

The resulting spin-independent cross sections are plotted in Fig. \ref{fig:SigmaXN} for the parameters that satisfy the relic abundance constraint (\ref{eq:planck}) and lie within the ranges (\ref{eq:ranges}). We see that there are significant regions in parameter space allowed by the LUX  constraint \cite{Akerib:2013tjd} (the ATLAS constraint in the figure is discussed below). In figure \ref{fig:EpsSigma} we plot the effective couplings $\epsilon_H$, $\epsilon_Z$ defined in (\ref{eq:eps}) that satisfy both LUX and PLANCK results.

 For the neutrino portal the spin-dependent cross sections are generated by the axial $Z$ couplings in (\ref{ZHex}) and are of the same order as the spin-independent ones. Super-K \cite{Choi:2015ara} and ICECUBE \cite{Aartsen:2012kia} have published the strongest limits on this  cross section, but these are still several orders of magnitude above the ones generated by (\ref{ZHex}) and put no significant constraints on the model parameters.

\subsection{Indirect detection}

Dark matter particles in the galactic halo have a finite probability to be elastically scattered by a nucleus and become subsequently trapped in the gravitational well of an astronomical object such as the Sun or Earth. These DM particles will undergo subsequent scatterings, until they thermalize and accumulate at the  core of the object~\cite{Feng:2010gw}. The accumulated DM particles in the inner core of the Sun or Earth can annihilate into SM particles that can be detected, among others, in astrophysical high energy neutrino experiments. 

In the neutrino-portal scenario neutrinos are the most abundant products of DM self-annihilation; the reaction occurs through the exchange of dark scalars into neutrinos (cf. fig ~\ref{fig:PPnn}). Given the small DM velocities the neutrino spectrum is essentially a delta function centered around $E_\nu\simeq \mfe$; the corresponding  spectrum detected on ground experiments  is given by \cite{Gould:1987ww,Lundberg:2004dn,Jungman:1995df}:
\beq
\frac{dN_\nu}{dE_\nu}\sim\frac{\Gamma_{\Psi\Psi\to\nu\bar{\nu}}}{4\pi R^2}\delta(E_\nu-m_\Psi)
\label{eq:nu-spectrum}
\eeq
where $\Gamma_{\Psi\Psi\to\nu\bar{\nu}}$ is the DM-DM annihilation rate, and $R$ is the distance from the neutrino source to the detector (Sun-Earth distance or Earth radius for neutrino annihilation in the Sun or Earth, respectively).  The annihilation rate is determined by the capture rate $C_\Psi $: when capture and annihilation processes reach equilibrium in a time-scale  much smaller than the age of the body (e.g. Sun or Earth), then $\Gamma_{\Psi\Psi \to \nu \bar\nu} \simeq C_\Psi/2$. The capture rate depends on the DM-nucleus cross sections, the DM velocity dispersion and the DM local density \cite{Jungman:1995df}, roughly $C_{\Psi}\propto \sigma_{\Psi\ncal} \rho^{DM}_{\rm local}$, where the first factor is generated by (\ref{ZHex}).

\begin{figure}[t]
$$\includegraphics[width=4.5in]{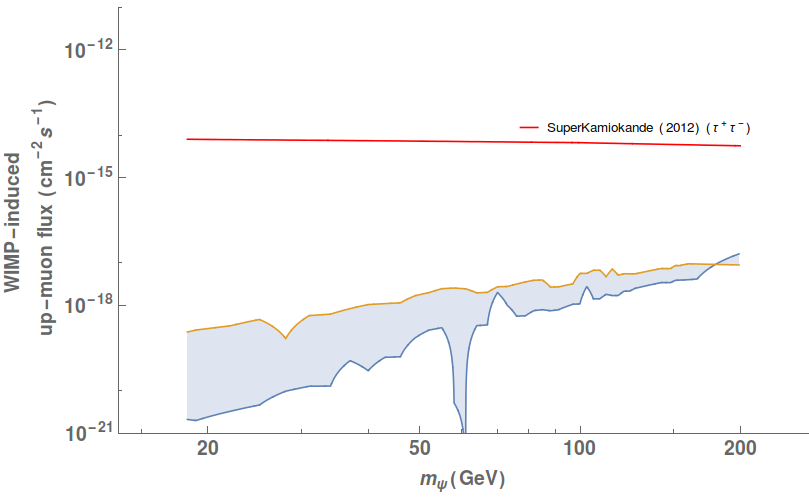}$$
\cprotect\caption{Blue shaded region shows the dark matter induced upward muon flux from the inner Solar core DM annihilation into neutrinos in the neutrino portal effective model. The red line is the Super-K limit over WIMP-induced (neutralino-like) upward muon flux from the Sun. }
\label{fig:UpMuonSK}
\end{figure}

Once produced the neutrinos will sometimes convert into muons as they move through the Earth, generating a upward-moving muon flux. We use the \verb|MicrOmegas| package to estimate both the neutrino and the upward muon fluxes. The calculation takes into account not only the dominating process (\ref{eq:nu-spectrum}), but all  $\Psi \Psi \rightarrow$ SM SM channels, and uses  the tabulated neutrino spectra functions \cite{Cirelli:2005gh}, taking into account effects induced by oscillation and attenuation processes. Figure \ref{fig:UpMuonSK} show the resulting upward muon flux for the parameter ranges (\ref{eq:ranges}) within the effective model, and, for comparison, the SuperKamiokande \cite{Tanaka:2011uf} WIMP-induced (neutralino-like) upward muon flux from inner Solar core dark matter annihilation into SM particles. The graph shows that within the neutrino portal scenario this neutrino flux is, in general, much smaller than what is expected in generic WIMP scenarios.

\begin{figure}[t]
$$\includegraphics[width=4.5in]{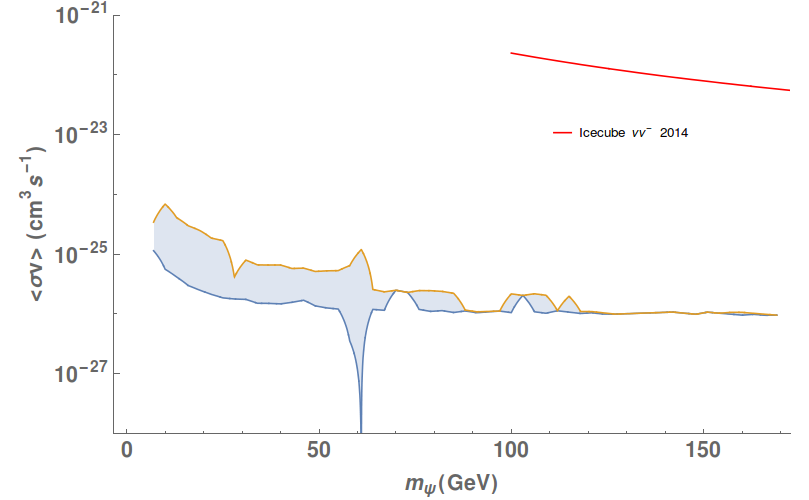}$$
\cprotect\caption{Blue shaded region shows the DM annihilation  cross section $\Psi\Psi\to\nu\nu$ within the neutrino-portal scenario. The red line is the ICECUBE limit (published results do not extend to masses below $ 100 \gev $) }
\label{fig:ICE3}
\end{figure}

The galactic halo is another source of DM annihilation products that may be accessible to indirect detection experiments. While the neutrino flux from DM annihilation captured in the Sun or Earth depends on the DM-nucleon cross section, the galactic neutrino flux depends on the self-annihilation cross section \cite{Yuksel:2007ac,Bertone:2004pz}.  ICECUBE experiment measures the characteristic anisotropic flux of highly energetic neutrinos for different annihilation channels, including direct annihilation into neutrinos \cite{Aartsen:2014hva}.; however, as shown in figure \ref{fig:ICE3},  the experimental sensitivity is still several orders of magnitude below the predictions of the model.

\subsection{Collider onstraints}

In this section we briefly cover the main existing collider experimental constraints on the effective-theory realization of the neutrino-portal scenario. These constraints are derived from the invisible decays of the $Z$ and Higgs generated by the reactions, $ Z \to \Psi\Psi $ and $ H \to \Psi \Psi $ (assuming they are kinematically allowed).
\bit
\item $Z$ invisible decay. For $m_Z > 2 \mfe $ 
\bea
\mathcal{BR}( Z\to \Psi \Psi) &\simeq& \frac{1.18 \times 10^{-9}}{\Lambda_{\rm TeV}^4}\sqrt{\left( 1-\frac{4\mfe^2}{m^2_Z} \right)} \cr
&\times& \left\{ \left( \left|\cvii\up{\phi|L}\right|^2 + \left| \cvii\up{\phi|R} \right|^2 \right)\left(1-2\frac{\mfe^2}{m^2_Z}\right)
+8\frac{\mfe^2}{m^2_Z}  \re\left( \cvii\up{\phi|R}{}^* \cvii\up{\phi|L} \right) \right\} \cr &&
\eea
where $\Lambda_{\rm TeV} = \Lambda/(1 \tev)$. Using (\ref{eq:ranges}) we find that $\bcal\rcal( Z\to \Psi \Psi)$ is at most of order $5\times 10^{-7}$, which is significantly below the error in the invisible branching ratio ${\cal BR}(Z \to {\rm inv} )=(20\pm 0.06)\%$. This process does not impose a significant restriction over the model parameters.

 \item $H$ invisible decay. There are two processes that contribute: $ H \to \Psi\Psi,~ \Phi \Psi \nu $ (the latter followed by the prompt decay $ \Phi \to \Psi\nu $).
 
For the first case, after a straightforward calculation we find, for $ m_H >2\mfe $,
\bea
\Gamma(H \to \Psi\Psi) &\simeq& \frac{(\cii v/\Lambda)^2 m_H }{2048 \pi^5}\left(1 - 4\frac{\mfe^2}{m^2_H} \right)^{3/2}
\eea
where we neglected a small contribution from the class VII operators. This then implies
\beq
\mathcal{BR}( H\to \Psi \Psi) \simeq 3\times 10^{-3}\left(\frac{\cii }{\Lambda_{\rm TeV}}\right)^2\left(1 - 4\frac{\mfe^2}{m^2_H} \right)^{3/2}
\eeq
where $ \Lambda_{\rm TeV} = \Lambda/(1 \tev) $. The direct search for Higgs invisible decays in the ATLAS experiment \cite{Aad:2014iia} sets an upper limit of $63\%$ at $90\%$ C.L. This  limit on $\mathcal{BR}$$( H\to \mathrm{inv})$ translates into an upper limit on the DM-nucleon spin-independent scattering and is shown in figure \ref{fig:SigmaXN}. There are no exclusion regions over the natural ranges of the effective couplings.
\eit

For the second case  we obtain
\beq
 \Gamma(H \to \nu \Psi \Phi) =  \frac{\ciii^2  m_H^3}{128 \pi^3 \Lambda^2}  \, A\left( \frac{m_H^2+ \mfe^2-\msc^2}{m_H^2+ \mfe^2}, \frac{m_H^2 - \mfe^2}{m_H^2 + \mfe^2} \right)
 \eeq
 where
 \bea
 A(b,c) &=& \frac{(4b + c + 1)(1-c)^2}{8(1+c)^2} \ln \left[ \frac{ b + \sqrt{b^2+c^2-1}}{\sqrt{1-c^2}} \right] 
 \cr && 
 - \frac{\sqrt{b^2+c^2-1}}{24(1+c)^3} \left[ 2 b^2(2-b - 2c) + (8+5b - 8c)(1-c^2) \right] \cr &&
 \eea
 The corresponding branching ratio is below 0.4\% for $ \mfe \ge 5 \gev $, and it again imposes no significant constraint on the model.

\section{Conclusions}
\label{sec:conclusions}
We discussed a general effective-theory description of a multi-component dark sector and its interactions with the SM generated by the exchange of heavy mediators. We obtained all effective interactions generated by operators of dimension $ \le 6 $ for the case where all dark fields transform non-trivially under an unspecified symmetry group $ \gdm $ and are all singlets under the SM gauge group $ \gsm $, while all SM fields are assumed to be $ \gdm $ singlets.  We then specialized to the case where the underlying theory is weakly coupled and renormalizable, and when the mediators are neutral under $ \gdm \times \gsm $ and are either scalars or fermions. In this case we argued that only a relatively small number of the effective operators are relevant given the sensitivity of current and near-future experiments. 

We then consdiered in some detail the neutrino-portal scenario where only fermion mediators are present, and where the dark sector consists of fermions and scalars such that the lightest dark particle is a fermion. This scenario is characterized by having naturally suppressed couplings of the DM to all SM particles except the neutrinos, and has received little attention in the literature. We find that there is a wide region in parameter space allowed by the current experimental constraints (relic abundance, direct and indirect detection limits). In particular, the DM mass is unconstrained for a significant ranges of the remaining model parameters; still, an improvement in one order of magnitude in the experimental sensitivity would  exclude DM masses below $\sim m_H/2 $. We also considered the possible collider constraints on the model parameters and found them to be very weak. The cleanest signature of this scenario is the presence of a monochromatic neutrino line, from both the Sun and the halo, with energy equal to that of the DM mass, experimental sensitivity would have to be improved significantly before this can be probed.  

Though in this publication we concentrated on the effective theory approach to the interactions between the dark and standard sectors, the neutrino portal scenario is clearly easily studied using specific model realizations. We will present such a discussion in an upcoming publication.

\appendix

\section{The operators $ \ocal\up4$}
\label{sec:o4}
In this appendix we list the operators $ \ocal\up4$ referred to in table \ref{tab:ops}. Using gauge invariance we obtain
\beq
\ocal_{\rm SM}\up4 \in \left\{ |\phi|^4,\,\square|\phi|^2,\,  \bar\psi \varphi \psi' ,\,B_\mn^2,\, (W_\mn^I)^2,\, (G_\mn^A)^2 \right\}
\eeq
where $\psi=\ell,\,q; ~ \psi'=u,\,d,\,e;~\varphi=\phi,\,\phit$ ($q,\,\ell$ denote the left-handed quark and lepton SM isodoublets respectively; $ u,\,d$ the right-handed quark isosinglets, and $e$ the corresponding right-handed lepton isosinglet; $ \phit = i \sigma_2 \phi^*$, where $ \sigma_2 $ is the usual Pauli matrix). The operators involving fermions correspond to the Yukawa terms in the SM. Operators such as $ \bar\psi i \not\!\!D\psi$ and $ |D\phi|^2 $ were not included because they generate redundant terms in the effective Lagrangian after application of the equivalence theorem~\cite{eqv-th}.

Similarly
\beq
\ocal_{\rm dark}\up4	\in \left\{ |\Phi|^4,\,\square|\Phi|^2,\, \Phi\bar\Psi  P_{L,R}\Psi ,\, X_\mn^2 \right\}
\eeq
and
\beq
\ocal_{{\rm dark}\,\mn}\up4 \in \left\{ \Phi^\dagger X_\mn \Phi,\, \Phi \bar\Psi \sigma_\mn P_{L,R} \Psi ,\, \bar\Psi \left( \gamma_\mu \dcal_\nu - \gamma_\nu \dcal_\mu \right) P_{L,R} \Psi \right\}
\eeq
where some terms may be absent for specific choices of the dark symmetry group $ \gdm $. Some possibilities such as $ \partial_\mu ( \Phi^\dagger \dcal_\nu \Phi) - ( \mu \leftrightarrow \nu ) $ have been eliminated by applying the equivalence theorem.

\section{Cross sections}
\label{sec:xs}

In this appendix we provide, for completeness, the expressions of the various cross sections used int he calculation of the relic abundance.
\paragraph{Neutrino final states}
\bea
\sigma(\Psi \Psi \to \nu \nu) &=& \frac{(v \ciii /\Lambda)^4}{256\pi s \beta_\Psi} \Biggl\{
\frac{1+2 x(1+x) - \beta_\Psi^2}{(1+x)^2 - \beta_\Psi^2} - (1-x) \, \re A_R \cr 
&& \quad + \inv4 \left(1 + \frac{\beta_\Psi^2}3 \right) \left( |A_L|^2 + |A_R|^2 \right) + \frac{\mfe^2}s \re(A_L^* A_R) \cr
&&+ \frac x\beta_\Psi \left( 1 + \frac x2 \, \re A_R + \frac{2 \mfe^2}{x s} \, \re A_L\right) \ln \left| \frac{1+x - \beta_\Psi}{1+x + \beta_\Psi} \right| \Biggr\}
\label{sigPPvv}
\eea
where
\beq
\beta_r= \sqrt{ 1 - \frac{4 m_r^2}s}\,, \quad x = \frac2s(\msc^2 - \mfe^2)\,, \quad
A_{L, R} = \left(\frac g{4 \pi c_{\rm w}} \right)^2 \frac{\cvii\up{\phi | L,R}}{\ciii^2} \frac s{s - m_Z^2 + i m_Z \Gamma_Z}
\label{eq:defs}
\eeq

\paragraph{Quark and charged lepton final states}

\bea
\sigma(\Psi\Psi\to u u) &=&\bar\sigma^u_H(s) +\bar\sigma_Z(s) \left[
3\acal y_\Psi(1-2y_u)-\frac{y_\Psi}{4}+1 \right.\cr
& +&\left.  \left(\frac{4}{3}\acal y_\Psi(1+2y_u)-\frac{4}{3}y_u\left(\frac{2}{3}y_\Psi+1\right)-\frac{y_\Psi}{9}+\frac{4}{9} \right) 2\sw^2(4 \sw^2-3)  \right]\\
&& \cr
&& \cr
\sigma(\Psi\Psi\to d d) &=&\bar\sigma^d_H(s) + \bar\sigma_Z(s) \left[
3\acal y_\Psi(1-2y_d)-\frac{y_\Psi}{4}+1 \right.\cr
& +&\left.  \left(\frac{4}{3}\acal y_\Psi(1+2y_d)-\frac{4}{3}y_d\left(2 y_\Psi-1\right)-\frac{y_\Psi}{9}+\frac{4}{9} \right) \sw^2(2 \sw^2-3)  \right]\\
&& \cr
&& \cr
\sigma(\Psi\Psi\to ll) &=&\bar\sigma^l_H(s) +16 \bar\sigma_Z(s) \left[
\acal y_\Psi(1-2y_l)-\frac{y_\Psi}{12}+\frac{1}{3}\right. \cr
&+&\left. \left(4\acal y_\Psi(1+2y_l)-4y_l( 2y_\Psi-1) -\frac{y_\Psi}{3}+\frac{4}{3}\right) \sw^2(2\sw^2-1)  \right]\\
\label{sigPPff}
\eea
where
 \bea
\bar\sigma_Z(s) &=&\frac{(v\ciii /\Lambda)^4(  |A_L |^2+ |A_R|^2)}{1024\pi s \beta_\Psi},\,  \qquad \acal = \frac{2\re\acal^{*}_L\acal_R}{ |A_L|^2 +| A_R |^2},\, \qquad y_i = \frac{m_i^2}{s}, \cr
\bar\sigma^f_H(s) &=& \frac{N_f\beta^3_{f} s }{64 \pi \beta_\Psi }\left(
\frac{m_{f}}{16\pi^2\Lambda^2}\right)^2 \left\{ \frac{ 4 \cii^2 \beta_\Psi^2 \Lambda^2 - \left[ \cvii\up{\phi | R}-\cvii\up{\phi | L} \right]^2 \mfe^2 }{(m^2_H-s)^2+m^2_H\Gamma^2_H}\right\}
\eea

\acknowledgments
This work was supported in part by a postdoctoral fellowship grant from UC Mexus. JW would like to thank F. del \'Aguila, B. Grzadkowski and A. Santamar\'\i a for insightful conversations.


\begin{thebibliography}{99}

\bibitem{Feng:2010gw} 
  See, for example, J.~L.~Feng,
  Ann.\ Rev.\ Astron.\ Astrophys.\  {\bf 48}, 495 (2010)
  [arXiv:1003.0904 [astro-ph.CO]].

\bibitem[Tegmark (2003)]{Tegmark} 
  M.~Tegmark {\it et al.}  [SDSS Collaboration],
  Phys.\ Rev.\ D {\bf 69}, 103501 (2004)
  [astro-ph/0310723].

\bibitem[Perlmutter (1998)]{Perlmutter} 
  S.~Perlmutter {\it et al.}  [Supernova Cosmology Project Collaboration],
  Astrophys.\ J.\  {\bf 517}, 565 (1999)
  [astro-ph/9812133].



\bibitem[Tonry (2003)]{Tonry} 
  J.~L.~Tonry {\it et al.}  [Supernova Search Team Collaboration],
  Astrophys.\ J.\  {\bf 594}, 1 (2003)
  [astro-ph/0305008].

\bibitem[Riess (1998)]{Riess} 
  A.~G.~Riess {\it et al.}  [Supernova Search Team Collaboration],
  Astron.\ J.\  {\bf 116}, 1009 (1998)
  [astro-ph/9805201].


\bibitem{Choi:2015ara} 
  K.~Choi {\it et al.}  [Super-Kamiokande Collaboration],
  Phys.\ Rev.\ Lett.\  {\bf 114}, no. 14, 141301 (2015)
  [arXiv:1503.04858 [hep-ex]].

\bibitem{Aartsen:2012kia} 
  M.~G.~Aartsen {\it et al.}  [IceCube Collaboration],
  Phys.\ Rev.\ Lett.\  {\bf 110}, no. 13, 131302 (2013)
  [arXiv:1212.4097 [astro-ph.HE]].


\bibitem{Akerib:2013tjd} 
  D.~S.~Akerib {\it et al.}  [LUX Collaboration],
  Phys.\ Rev.\ Lett.\  {\bf 112}, 091303 (2014)
  [arXiv:1310.8214 [astro-ph.CO]].

\bibitem{Aprile:2012nq} 
  E.~Aprile {\it et al.}  [XENON100 Collaboration],
  Phys.\ Rev.\ Lett.\  {\bf 109}, 181301 (2012)
  [arXiv:1207.5988 [astro-ph.CO]].

\bibitem{Ade:2013zuv} 
  P.~A.~R.~Ade {\it et al.}  [Planck Collaboration],
  Astron.\ Astrophys.\  {\bf 571}, A16 (2014)
  [arXiv:1303.5076 [astro-ph.CO]].

\bibitem{Goodman:2010ku}
  J.~Goodman, M.~Ibe, A.~Rajaraman, W.~Shepherd, T.~M.~P.~Tait and H.~B.~Yu,
  Phys.\ Rev.\ D {\bf 82} (2010) 116010
  [arXiv:1008.1783 [hep-ph]].
\bibitem{Belanger:2008sj}
  G.~Belanger, F.~Boudjema, A.~Pukhov and A.~Semenov,
  Comput.\ Phys.\ Commun.\  {\bf 180} (2009) 747
  [arXiv:0803.2360 [hep-ph]].
\bibitem{Duch:2014xda}
  M.~Duch, B.~Grzadkowski and J.~Wudka,
  JHEP {\bf 1505} (2015) 116
  [arXiv:1412.0520 [hep-ph]].
  
\bibitem{Appelquist:1974tg}
  T.~Appelquist and J.~Carazzone,
  Phys.\ Rev.\ D {\bf 11} (1975) 2856.
 
  
\bibitem{Drozd:2011aa}
  A.~Drozd, B.~Grzadkowski and J.~Wudka,
  JHEP {\bf 1204} (2012) 006
   [JHEP {\bf 1411} (2014) 130]
  [arXiv:1112.2582 [hep-ph]].
\bibitem{Bhattacharya:2013hva}
  S.~Bhattacharya, A.~Drozd, B.~Grzadkowski and J.~Wudka,
  JHEP {\bf 1310} (2013) 158
  [arXiv:1309.2986 [hep-ph]].
\bibitem{Kumar:2013tra}
  H.~Kumar and S.~Alam,
  arXiv:1307.7469 [astro-ph.CO].
  
  
\bibitem{Silveira:1985rk} 
  V.~Silveira and A.~Zee,
  Phys.\ Lett.\ B {\bf 161}, 136 (1985).
\bibitem{McDonald:1993ex} 
  J.~McDonald,
  Phys.\ Rev.\ D {\bf 50}, 3637 (1994)
  [hep-ph/0702143 [HEP-PH]].
\bibitem{Burgess:2000yq} 
  C.~P.~Burgess, M.~Pospelov and T.~ter Veldhuis,
  Nucl.\ Phys.\ B {\bf 619}, 709 (2001)
  [hep-ph/0011335].
\bibitem{Bento:2000ah} 
  M.~C.~Bento, O.~Bertolami, R.~Rosenfeld and L.~Teodoro,
  Phys.\ Rev.\ D {\bf 62}, 041302 (2000)
  [astro-ph/0003350].
\bibitem{Holz:2001cb} 
  D.~E.~Holz and A.~Zee,
  Phys.\ Lett.\ B {\bf 517}, 239 (2001)
  [hep-ph/0105284].
\bibitem{LopezHonorez:2006gr} 
  L.~Lopez Honorez, E.~Nezri, J.~F.~Oliver and M.~H.~G.~Tytgat,
  JCAP {\bf 0702}, 028 (2007)
  [hep-ph/0612275].


\bibitem{Falkowski:2011xh} 
  A.~Falkowski, J.~T.~Ruderman and T.~Volansky,
  JHEP {\bf 1105}, 106 (2011)
  [arXiv:1101.4936 [hep-ph]].
\bibitem{Baek:2013qwa} 
  S.~Baek, P.~Ko and W.~I.~Park,
  JHEP {\bf 1307}, 013 (2013)
  [arXiv:1303.4280 [hep-ph]].
\bibitem{Cosme:2005sb} 
  N.~Cosme, L.~Lopez Honorez and M.~H.~G.~Tytgat,
  Phys.\ Rev.\ D {\bf 72}, 043505 (2005)
  [hep-ph/0506320].
\bibitem{An:2009vq} 
  H.~An, S.~L.~Chen, R.~N.~Mohapatra and Y.~Zhang,
  JHEP {\bf 1003}, 124 (2010)
  [arXiv:0911.4463 [hep-ph]].


\bibitem{Kolb:1990vq} 
  E.~W.~Kolb and M.~S.~Turner,
  Front.\ Phys.\  {\bf 69}, 1 (1990).


\bibitem{Belanger:2013oya} 
  G.~Belanger, F.~Boudjema, A.~Pukhov and A.~Semenov,
  Comput.\ Phys.\ Commun.\  {\bf 185}, 960 (2014)
  [arXiv:1305.0237 [hep-ph]].

\bibitem{Belyaev:2012qa} 
  A.~Belyaev, N.~D.~Christensen and A.~Pukhov,
  Comput.\ Phys.\ Commun.\  {\bf 184}, 1729 (2013)
  [arXiv:1207.6082 [hep-ph]].

\bibitem{Alloul:2013bka} 
  A.~Alloul, N.~D.~Christensen, C.~Degrande, C.~Duhr and B.~Fuks,
  Comput.\ Phys.\ Commun.\  {\bf 185}, 2250 (2014)
  [arXiv:1310.1921 [hep-ph]].



\bibitem{Aad:2014iia} 
  G.~Aad {\it et al.}  [ATLAS Collaboration],
  Phys.\ Rev.\ Lett.\  {\bf 112}, 201802 (2014)
  [arXiv:1402.3244 [hep-ex]].


\bibitem{Gould:1987ww} 
  A.~Gould,
  Astrophys.\ J.\  {\bf 328}, 919 (1988).

\bibitem{Lundberg:2004dn} 
  J.~Lundberg and J.~Edsjo,
  Phys.\ Rev.\ D {\bf 69}, 123505 (2004)
  [astro-ph/0401113].

\bibitem{Jungman:1995df} 
  G.~Jungman, M.~Kamionkowski and K.~Griest,
  Phys.\ Rept.\  {\bf 267}, 195 (1996)
  [hep-ph/9506380].

\bibitem{Cirelli:2005gh} 
  M.~Cirelli, N.~Fornengo, T.~Montaruli, I.~A.~Sokalski, A.~Strumia and F.~Vissani,
  Nucl.\ Phys.\ B {\bf 727}, 99 (2005)
  [Nucl.\ Phys.\ B {\bf 790}, 338 (2008)]
  [hep-ph/0506298].

\bibitem{Tanaka:2011uf} 
  T.~Tanaka {\it et al.}  [Super-Kamiokande Collaboration],
  Astrophys.\ J.\  {\bf 742}, 78 (2011)
  [arXiv:1108.3384 [astro-ph.HE]].

\bibitem{Yuksel:2007ac} 
  H.~Yuksel, S.~Horiuchi, J.~F.~Beacom and S.~Ando,
  Phys.\ Rev.\ D {\bf 76}, 123506 (2007)
  [arXiv:0707.0196 [astro-ph]].


\bibitem{Bertone:2004pz} 
  G.~Bertone, D.~Hooper and J.~Silk,
  Phys.\ Rept.\  {\bf 405}, 279 (2005)
  [hep-ph/0404175].

\bibitem{Aartsen:2014hva} 
  M.~G.~Aartsen {\it et al.}  [IceCube Collaboration],
  Eur.\ Phys.\ J.\ C {\bf 75}, no. 1, 20 (2015)
  [arXiv:1406.6868 [astro-ph.HE]].

  
    \bibitem{eqv-th}
  H.~Georgi,
  ``On-shell effective field theory,''
  Nucl.\ Phys.\ B {\bf 361} (1991) 339;
 J.~Wudka,
  ``Electroweak effective Lagrangians,''
  Int.\ J.\ Mod.\ Phys.\ A {\bf 9}, 2301 (1994)
  [hep-ph/9406205];
  C.~Arzt,
  ``Reduced effective Lagrangians,''
  Phys.\ Lett.\ B {\bf 342}, 189 (1995)
  [hep-ph/9304230].


\end{thebibliography}
\end{document}